\documentclass{ws-mpla}
\usepackage{amsmath,amssymb}

\begin{document}

\markboth{Y. S. Duan, L. D. Zhang and Y. X. Liu} {Self-dual
Vortices in the Abelian Chern-Simons Model with Two Complex Scalar
Fields}

\catchline{}{}{}{}{}

\title{SELF-DUAL VORTICES IN THE ABELIAN CHERN-SIMONS
MODEL WITH TWO COMPLEX SCALAR FIELDS}

\author{
 Yi-Shi Duan,
 Li-Da Zhang\footnote{Corresponding author.
                       Email: zhangld04@lzu.cn},
 Yu-Xiao Liu}
\address{Institute of Theoretical Physics, Lanzhou University,\\
   Lanzhou 730000, P. R. China}

\maketitle

\begin{abstract}
Making use of $\phi$-mapping topological current method, we
discuss the self-dual vortices in the Abelian Chern-Simons model
with two complex scalar fields. For each scalar field, an exact
nontrivial equation with a topological term which is missing in
many references is derived analytically. The general angular
momentum is obtained. The magnetic flux which relates the two
scalar fields is calculated. Furthermore, we investigate the
vortex evolution processes, and find that because of the present
of the vortex molecule, these evolution processes is more
complicated than the vortex evolution processes in the
corresponding single scalar field model.

\keywords{Self-dual Vortices; Abelian Chern-Simons Model.}
\end{abstract}

\ccode{PACS numbers: 11.15.-q, 02.40.Pc, 47.32.Cc}

\section{INTRODUCTION}
The (2+1)-dimensional physics system presents many interesting
surprises, both experimentally and theoretically. For example, in
last three decades, it has been realized that while in three and
higher space dimensions the particles must be either bosons or
fermions, in two space dimensions the particles can have any
fractional spin and can satisfy any fractional statistics which
interpolates between the familiar two.\cite{Khare1997} The
particles obeying such statistics are called as anyons, whose
existence in real physics system have been proved by the
fractional quantum Hall
effect\cite{TsuiPRL198248,LaughlinPRL198350} where anyons act as
fractionally charged excitations of the imcompressible quantum
fluid.

In (2+1)-dimensional spacetime, a new type of gauge theory named
Chern-Simons theory which is different from Maxwell theory plays a
very significant role in many
aspects.\cite{Khare1997,Dunne9902115} One of the most remarkable
properties of the Chern-Simons action is that it does not depends
on the metric tensor, and always has the same form no matter the
spacetime is flat or not. Hence, the Chern-Simons theory is an
important example of the topological field
theory.\cite{SchwarzLMP19782,WittenCMP1989121,BosPLB1989223,BirminghamPR1991209}
There are both topological and nontopological self-dual vortex
solutions in the Chern-Simons
model.\cite{HongPRL199064,JackiwPRL199064} And topology is
indispensable even in the situation where only the nontopological
vortex solutions are present.

On the other hand, for over four decades there has been a wide
interest in the condensed-matter systems with several coexisting
Bose condensates.\cite{SuhlPRL19593} Recent
research\cite{BabaevPRB200265} has revealed that a charged
two-condensate Ginzburg-Landau model can be mapped onto a version
of the nonlinear $O(3)$ $\sigma$ model. This leads to many efforts
to study the topological property behind the two-condensate
system.\cite{BabaevPRL200288,BabaevPRL200289,JiangPRB200470,DuanPRB200674}

In light of $\phi$-mapping topological current
method,\cite{DuanJMP200041} the present paper investigates the
self-dual vortices in the Abelian Chern-Simons model with two
complex scalar fields, which is a kind of (2+1)-dimensional field
theory generalization of the two-condensate system. To be
self-contained basically, in Sec. \ref{SecMap}, we briefly review
the $\phi$-mapping topological current method. In Sec.
\ref{SecModel}, we introduce the model to be considered, and
derive a set of self-dual equations. In Sec. \ref{SecVortex}, we
find a nontrivial equation with a topological term for each scalar
field, which is the development of our previous
work,\cite{Wang0508104,WangMPLA200520,LiuNPB2007} and also obtain
an equation which relates the two scalar fields by their
topological terms. Besides, the angular momentum and the magnetic
flux of the system are calculated. In Sec. \ref{SecBranch}, we
briefly consider the evolution processes of the vortices, and find
that because of the present of the vortex molecule, the detailed
evolution processes of the vortices in the double scalar field
model is more complicated than the vortex evolution processes in
the corresponding single scalar field model, which have been
investigated in Ref. \refcite{FuPRD2000045004} in detail.

\section{A BRIEF REVIEW OF $\phi$-MAPPING TOPOLOGICAL CURRENT
METHOD} \label{SecMap}

 Because the $\phi$-mapping topological
current method is unfamiliar to most of the readers, we will
briefly review it in the following. Consider a D-dimensional
smooth manifold with metric tensor $g_{\mu\nu}$ and local
coordinates $x^{\mu}$. Define a map $\Phi$:
\begin{equation}\label{map}
\Phi^{a}=\Phi^{a}(x^{\mu})~~~(a=1,\cdots,d<D),
\end{equation}
and introduce the direction unit field of $\Phi^{a}$:
\begin{equation}\label{na}
N^{a}=\frac{\Phi^{a}}{\|\Phi\|}~~~(\|\Phi\|=\sqrt{\Phi^{a}\Phi^{a}}).
\end{equation}
Using $N^{a}$, we can construct a topological tensor current:
\begin{equation}\label{jmn0}
j^{\mu\cdots\nu}\sim\frac{1}{\sqrt{detg_{\mu\nu}}}~\epsilon^{\mu\cdots\nu\lambda\cdots\rho}
 \epsilon^{a\cdots b}\partial_{\lambda}N^{a}\cdots\partial_{\rho}N^{b}.
\end{equation}
It is not difficult to find that $j^{\mu\cdots\nu}$ is completely
antisymmetric and identically conserved.
 Defined the Jacobian tensor $D^{\mu\cdots\nu}(\frac{\Phi}{x})=\frac{1}{d!}
 ~\epsilon^{\mu\cdots\nu\lambda\cdots\rho}
 \epsilon^{a\cdots
 b}\partial_{\lambda}\Phi^{a}\cdots\partial_{\rho}\Phi^{b}$, and the $\phi$-space
Green function $G_{d}(\|\Phi\|)=\left\{%
\begin{array}{ll}
    \frac{1}{\|\Phi\|^{d-2}}, & {d>2;} \\
    \ln\|\Phi\|, & {d=2.} \\
\end{array}%
\right.    $
 Then using
 $\partial_{\mu}N^{a}=\frac{1}{\|\Phi\|}\partial_{\mu}\Phi^{a}+\Phi^{a}\partial_{\mu}
 \frac{1}{\|\Phi\|}$ and the Green function relation $\frac{\partial}{\partial\Phi^{a}}
\frac{\partial}{\partial\Phi^{a}}G_{d}(\|\Phi\|)\sim\delta^{d}(\Phi)$
we can find
\begin{equation}\label{jmn01}
j^{\mu\cdots\nu}\sim\frac{1}{\sqrt{detg_{\mu\nu}}}~\delta^{d}(\Phi)
D^{\mu\cdots\nu}(\frac{\Phi}{x}).
\end{equation}
The $\delta$-function in Eq. (\ref{jmn01}) implies that
$j^{\mu\cdots\nu}\neq0$ only at the points where $\Phi^{a}=0$.
Generally, these points would correspond to the submanifolds where
the topological defects are located. Denoting the $k$-th
above-mentioned submanifold by $M_{k}$, and defining a
corresponding normal submanifold $N_{k}$ which is spanned by the
parameter $v^{A}$ $(A=1,\cdots,k)$ with the metric tensor
$g_{AB}$, we can get the only intersection point of $M_{k}$ and
$N_{k}$ denoted by $p_{k}$. According to the implicit function
theorem,\cite{Goursat1904} at the regular points of $\Phi^{a}$,
there exists the only solution to the equations $\Phi^{a}=0$, and
Eq. (\ref{jmn01}) can be expanded as
\begin{equation}\label{jmn02}
j^{\mu\cdots\nu}\sim\frac{1}{\sqrt{detg_{\mu\nu}}}~\sum_{k}
\frac{W_{k}\sqrt{detg_{AB}}}{\left.D(\frac{\Phi}{v})\right|_{p_{k}}}~\delta^{d}(M_{k})
D^{\mu\cdots\nu}(\frac{\Phi}{x}),
\end{equation}
where the Jacobian $D(\frac{\Phi}{v})=\frac{1}{d!}
 ~\epsilon^{A\cdots B}
 \epsilon^{a\cdots
 b}\partial_{A}\Phi^{a}\cdots\partial_{B}\Phi^{b}$,
 $\delta^{d}(M_{k})$ is the $\delta$-function on the submanifold $M_{k}$,
  and $W_{k}$ denotes the winding number of the
 $k$-th topological defect. Some important physical quantities
 concerning the topology, say magnetic flux, can be calculated from Eq.
 (\ref{jmn02}). By virtue of the implicit function theorem,\cite{Goursat1904}
 the irregular points of $\Phi^{a}$ correspond to the branch
 points of the topological current $j^{\mu\cdots\nu}$, and the
 branch processes of $j^{\mu\cdots\nu}$ occur at these very
 points. These branch processes can describe various evolutions of
 the topological defects, and the total topological charge of the
 system will keep unchange during these evolutions since $j^{\mu\cdots\nu}$
 is a conserved current.

\section{THE MODEL}
\label{SecModel}

We consider the (2+1)-dimensional Chern-Simons system described by
the Lagrangian
\begin{equation}\label{L}
{\cal L}=\frac{1}{2} \kappa\varepsilon^{\mu\nu\rho} a_\mu
\partial_{\nu} a_\rho+|D_\mu\psi|^2+|D_\mu\phi|^2-U(\psi,\phi),
\end{equation}
where $D_\mu\psi=(\partial_{\mu}-ie_1a_\mu)\psi,$ $
D_\mu\phi=(\partial_{\mu}-ie_2a_\mu)\phi$, and
 \begin{eqnarray}
  \label{U}
U(\psi, \phi)=\frac{1}{\kappa^2} \left(e_1^2|\psi|^2
+e_2^2|\phi|^2\right)\left(e_1|\psi|^2+ e_2|\phi|^2-v^2\right)^2.
\end{eqnarray}
Here, coupling $e_1, e_2>0$, $\psi$ and $\phi$ are complex scalar
fields, $a_{\mu}$ denotes the Chern-Simons gauge field.
The specific form of the potential (\ref{U}) gives rise to a
self-dual system.

The variation of the action with respect to $a_{0}$ leads to a
Gauss law:
  \begin{equation}\label{gau1}
\kappa f_{12}=e_1J^0_\psi+e_2J^0_\phi,   \\
\end{equation}
where $f_{\mu\nu}=\partial_{\mu} a_\nu-\partial_{\nu} a_\mu$,
$J^\mu_\psi=-i[\psi^*D^\mu\psi-(D^\mu\psi)^*\psi]$ and
$J^\mu_\phi=-i[\phi^*D^\mu\phi-(D^\mu\phi)^*\phi]$.
 For finite-energy configurations, the magnetic flux
 $\Phi=\int\! d^2x f_{12}$ and charges $Q_\psi=\int\! d^2x J^0_\psi$
and $Q_\phi=\int\! d^2x J^0_\phi$ are related by Eq. (\ref{gau1}),
namely
\begin{equation} \label{gau2}
\kappa\Phi=e_1Q_\psi+e_2Q_\phi.
\end{equation}

Set $i, j=1, 2$, the energy function is
 \begin{equation}\label{E1}
E=\int\!d^2x[|D_0\psi|^2+|D_0\phi|^2+|D_i\psi|^2+|D_i\phi|^2+U].
\end{equation}
Employing the Gauss laws and integrating by parts, one obtains the
bound
 \begin{equation} \label{E2}
   E\geq v^2|\Phi|.
   \end{equation}
This bound is saturated by the fields satisfying the following
self-dual equations:
 \begin{eqnarray}
&&(D_1\pm iD_2)\psi = (D_1\pm i D_2)\phi=0,\nonumber\\
&&D_0\psi\pm i\frac{e_1}{\kappa}\psi(e_1|\psi|^2+ e_2|\phi|^2
-v^2)=0, \label{sd}\\
&&D_0\phi\pm i\frac{e_2}{\kappa}\phi(e_1|\psi|^2+ e_2|\phi|^2
-v^2)=0.\nonumber
\end{eqnarray}

\section{STATIC VORTICES}\label{SecVortex}

From the potential (\ref{U}), it is obvious that there is a
symmetric phase as well as an asymmetric or broken phase. In the
asymmetric phase, static finite energy solutions require both
scalar fields $\psi$ and $\phi$ to be at most a pure phase on the
circle at infinity, i.e. the vacuum manifold for both fields
$\psi$ and $\phi$ is $U(1)\simeq S^{1}$. So the first homotopy
group $\pi_{1}(U(1))=Z$ provides the topological guarantee for the
existence of the vortex solutions. In the symmetric phase,
nontopological static soliton solutions are possible, which make
the situation more complicated than the self-dual Maxwell system.

 For static case, from Eq. (\ref{gau1}) we obtain
\begin{equation} \label{a0}
a_0=-\frac{\kappa f_{12}}{2(e_1^2|\psi|^2+e_2^2|\phi|^2)}.
\end{equation}
From Eq. (\ref{sd}), we have
\begin{equation} \label{ai}
e_1a_{i}=\pm\epsilon^{0ij}\partial_{j}\ln|\psi|-\frac{i}{2}
(\frac{\psi^*}{|\psi|}\partial_i\frac{\psi}{|\psi|}
-\frac{\psi}{|\psi|}\partial_i\frac{\psi^*}{|\psi|}),
\end{equation}
and furthermore
\begin{equation} \label{f120}
 e_1f_{12}=e_1\epsilon^{0ij}\partial_{i}a_{j}
 =\mp\partial_{i}\partial_{i}\ln|\psi|
  -i\epsilon^{0ij}\partial_{i}\frac{\psi^*}{|\psi|}
                  \partial_{j}\frac{\psi}{|\psi|}.
\end{equation}
At this point, expressing $\psi$ as $\psi=\psi^1+i\psi^2$, we can
write the second term on the right-hand side of Eq. (\ref{f120})
as $\epsilon^{0ij}\epsilon_{ab}\partial_i
(\psi^a/|\psi|)\partial_j(\psi^b/|\psi|).$ This expression has an
exact form of topological current (\ref{jmn0}). So, using the
$\phi$-mapping method, we can rewrite Eq. (\ref{f120}) as
\begin{equation} \label{f121}
e_1f_{12}=e_1\epsilon^{0ij}\partial_{i}a_{j}=\mp\bigtriangledown^{2}\ln|\psi|
+2\pi\delta(\psi)D(\frac{\psi}{x}),
\end{equation}
where the Jacobian
$D(\frac{\psi}{x})=\frac{1}{2}\epsilon^{0ij}\epsilon_{ab}
\partial_i\psi^{a}\partial_j\psi^{b}$ and $\delta(\psi)$ is the
$\delta$-function in $\psi$ space. The presence of $\delta(\psi)$
in Eq. (\ref{f121}) identifies the existence of the vortex
solutions in model (\ref{L}), and implies that the cores of the
vortices locate at the zeros of $\psi$. To be specific, we denote
the isolated zeros of $\psi$ by $\vec{z}_{r}(r=1,\ldots,m)$ and
the isolated zeros of $\phi$ by $\vec{z}_{s}(s=1,\ldots,n)$.
 According to the implicit function theorem,\cite{Goursat1904} when
the zeros $\vec{z}_{r}$ are the regular points of $\psi$, i.e.
when
\begin{equation}\label{D1}
\left.D(\frac{\psi}{x})\right|_{\vec{z}_{r}}\neq 0,
\end{equation}
there exists one and only one continuous vortex solution to the
equations $\psi^{a}=0  (a=1,2)$. Furthermore, under the condition
(\ref{D1}), we can express $\delta(\psi)$
as\cite{Schouten1951,DuanJHEP20040402}
\begin{equation}\label{del1}
\delta(\psi)=\sum_{r=1}^m\frac{W_{r}}{D(\frac{\psi}{x})|_{\vec{z}_{r}}}
\delta(\vec{x_{r}}-\vec{z}_{r}),
\end{equation}
where $W_{r}$ is the winding number of the $r$-th isolated zero of
$\psi$. If we consider the case with $\psi$ vanishing at infinity
in the following way
\begin{equation} \label{bc1}
|\psi|\rightarrow\frac{1}{r^\alpha}\hspace{7mm}\mbox{as
}r\rightarrow\infty,
\end{equation}
where $\alpha>0$. Then
\begin{equation} \label{delD}
\delta(\psi)D(\frac{\psi}{x})\rightarrow0\hspace{7mm}\mbox{as
}r\rightarrow\infty.
\end{equation}
Therefore Eq. (\ref{f121}) can generally be rewritten as
 \begin{equation} \label{f122}
e_1f_{12}=\mp\bigtriangledown^{2}\ln|\psi| +2\pi\sum_{r=1}^mW_{r}
\delta(\vec{x}-\vec{z}_{r}),
\end{equation}
even though $\psi$ vanishes at infinity in some kind of specific
way.

Using Eqs. (\ref{sd}), (\ref{a0}) and (\ref{f122}), we find a
 nontrivial equation to be satisfied by $\psi$:
\begin{equation} \label{sde1}
\nabla^2\ln|\psi|^2-\frac{4e_1}{\kappa^2}(e_1^2|\psi|^2+e_2^2|\phi|^2)\left(e_1|\psi|^2+
       e_2|\phi|^2-v^2\right)=\pm4\pi
\sum_{r=1}^mW_{r} \delta(\vec{x}-\vec{z}_{r}).
\end{equation}
The above nontrivial equation includes a topological term on its
right-hand side, which is missing in many references. Under
$e_1\leftrightarrow e_2$, $\psi\leftrightarrow \phi$,
$\vec{z}_{r}\leftrightarrow \vec{z_{s}}$ and $W_r\leftrightarrow
W_s$ (the winding number of the $s$-th isolated zero of $\phi$),
all conclusions and expressions we obtain above still hold for
$\phi$. Especially, corresponding to Eq. (\ref{sde1}), we have
\begin{equation} \label{sde2}
\nabla^2\ln|\phi|^2-\frac{4e_2}{\kappa^2}(e_1^2|\psi|^2+e_2^2
 |\phi|^2)\left(e_1|\psi|^2+e_2|\phi|^2-v^2\right)=\pm4\pi
 \sum_{s=1}^nW_{s} \delta(\vec{x}-\vec{z_{s}}).
\end{equation}
From Eqs. (\ref{sde1}) and (\ref{sde2}), we find the following
equation for $\psi$ and $\phi$:
\begin{equation} \label{sde3}
\nabla^2\ln\frac{|\psi|^{e_2}}{|\phi|^{e_1}}=\pm2\pi\left[e_2
\sum_{r=1}^mW_{r} \delta(\vec{x}-\vec{z}_{r})-e_1\sum_{s=1}^nW_{s}
\delta(\vec{x}-\vec{z_{s}})\right].
\end{equation}
Obviously, the above equation relates $\psi$ and $\phi$ by their
topological terms, and displays the topological relation between
them.

 A characteristic feature of Chern-Simons vortices is that they
 carry nonzero angular momentum which is given by
\begin{equation}  \label{J1}
 J=-\int\! d^2x
\epsilon^{ij}x^i[(D_0\psi)^*D_j\psi
    +(D_0\phi)^*D_j\phi+\mbox{c.c.}].
\end{equation}
Furthermore, from Eq. (\ref{ai}) and its $\phi$ counterpart as
well as Eq. (\ref{sd}), Eq. (\ref{J1}) becomes
\begin{eqnarray}\label{J2}
J&&=-2~\!\!\!\int\! d^2x\epsilon^{ij}x^ia_0
    \left[|\psi|^2(e_1^2a_j-e_1\epsilon^{ab}\hat{\psi}^{a}
       \partial_{j}\hat{\psi}^{b})+|\phi|^2(e_2^2a_j-e_2\epsilon^{ab}\hat{\phi}^{a}
          \partial_{j}\hat{\phi}^{b})\right]\nonumber\\
 &&=\frac{1}{\kappa}~\!\!\!\int\! d^2xx^i(e_1|\psi|^2+e_2|\phi|^2-v^2)
    \partial_{i}(e_1|\psi|^2+e_2|\phi|^2)  \nonumber\\
 &&=\left\{%
\begin{array}{ll}
    -\frac{1}{\kappa}~\!\!\!\int\!
     d^2x(e_1|\psi|^2+e_2|\phi|^2-v^2)^{2}, & \hbox{asymmetric phase;} \\
    -\frac{1}{\kappa}~\!\!\!\int\!
 d^2x(e_1|\psi|^2+e_2|\phi|^2)(e_1|\psi|^2+e_2|\phi|^2-2v^2), & \hbox{symmetric phase.} \\
\end{array}%
\right.   
\end{eqnarray}
    This conclusion is exactly the generalization of the angular momentum of
single complex scalar Chern-Simons vortices.\cite{JackiwPRD199042}

In the following, we concentrate on asymmetric phase and assume
\begin{equation} \label{vv1}
\left(\begin{array}{c}\psi\\
\phi\end{array}\right) \rightarrow\left(\begin{array}{c}\psi_0\\
\phi_0\end{array}\right)\hspace{7mm}\mbox{as }r\rightarrow\infty,
\end{equation}
where $\psi_0$ and $\phi_0$  satisfy
$e_1|\psi_0|^2+e_2|\phi_0|^2=u^2$. Because different choices for
the vacuum values are inequivalent, we need consider the case with
$\psi_0\phi_0\neq 0$ and the case with one of the vacuum values is
zero respectively.

In the case with $\psi_0\phi_0\neq 0$, from Eq. (\ref{f122}) and
its $\phi$ counterpart, the flux is given by
\begin{equation} \label{phi1}
\Phi=\int\! d^2x
f_{12}=\frac{2\pi\sum_{r=1}^mW_{r}}{e_{1}}=\frac{2\pi\sum_{s=1}^nW^{'}_{s}}{e_{2}},
\end{equation}
which implies that the regular vortex solutions may exist only if
$e_{1}/e_{2}$ is rational and equals to the ratio of the total
winding number of zeros of $\psi$ and $\phi$. On the other hand,
in the case with one of the scalar fields, say $\psi$, vanishes at
infinity in the way of Eq. (\ref{bc1}), from Eq. (\ref{f122}) and
its $\phi$ counterpart the flux will be
\begin{equation} \label{phi2}
\Phi=\frac{2\pi(\sum_{r=1}^mW_{r}\pm\alpha)}{e_1}=\frac{2\pi\sum_{s=1}^nW_{s}}{e_2},
\end{equation}
where $+\alpha(-\alpha)$ is for $\sum_{r=1}^mW_{r}>0(<0).$

For the nontopological soliton solutions
 allowed in symmetric phase, the situation is different from the
 topological soliton solutions in some
 aspects. For example, suppose that for $\alpha>0$ and $\beta>0$
\begin{eqnarray}\label{vv2}
|\psi|\rightarrow\frac{1}{r^\alpha}\hspace{7mm}\mbox{as
}r\rightarrow\infty,\nonumber\\
|\phi|\rightarrow\frac{1}{r^\beta}\hspace{7mm}\mbox{as
}r\rightarrow\infty.
\end{eqnarray}
 Then, as before, we can find the flux is given by
\begin{equation} \label{phi3}
\Phi=\frac{2\pi(\sum_{r=1}^mW_{r}\pm\alpha)}{e_{1}}
=\frac{2\pi(\sum_{s=1}^nW_{s}\pm\beta)}{e_{2}},
\end{equation}
where $+\alpha(-\alpha)$ is for $\sum_{r=1}^mW_{r}>0(<0)$ and
$+\beta(-\beta)$ is for $\sum_{s=1}^nW_{s}>0(<0)$.

\section{EVOLUTION PROCESSES OF VORTICES}
\label{SecBranch}

In Ref. \refcite{FuPRD2000045004}, the evolution of the
Chern-Simons vortices in the model with single complex scalar
field was discussed in detail. As for the present system, we can
directly generalize the results in Ref. \refcite{FuPRD2000045004}
to the vortex evolution processes concerning the individual scalar
field, say $\psi$. The key issue here is the evolution processes
concerning both $\psi$ and $\phi$.

For complex scalar field $\psi$, define topological current
\begin{equation}  \label{j}
j^{\mu}=\frac{1}{2\pi }\epsilon ^{\mu \nu \lambda}\epsilon
_{ab}\partial _\nu \frac{\psi^{a}}{|\psi|}\partial _\lambda
\frac{\psi^{b}}{|\psi|}.
\end{equation}
It is easy to see that $j^{\mu}$ is identically conserved, i.e.
$\partial_{\mu}j^{\mu}=0$. Similar to the case in Sec.
\ref{SecVortex}, we can obtain
$j^{\mu}=\delta(\psi)D^{\mu}(\frac{\psi}{x})$. According to the
implicit function theorem as before, when
\begin{equation}\label{D2}
\left.D(\frac{\psi}{x})\right|_{(\vec{z}_{r}, t)}\neq 0,
\end{equation}
 there exists one and only one
continuous vortex solution to the equations:
\begin{eqnarray}
\psi^{1}(x^{1},x^{2},t)&=&0,\label{eq1}\\
\psi^{2}(x^{1},x^{2},t)&=&0,\label{eq2}
\end{eqnarray}
which can be expressed as
\begin{equation}\label{sol1}
\vec{z}_{r}=\vec{z}_{r}(t),
\end{equation}
and $\delta(\psi)$ can be expressed as
\begin{equation}\label{del2}
\delta(\psi)=\sum_{r=1}^m\frac{W_{r}}{\left.D(\frac{\psi}{x})\right|_{(\vec{z}_{r},
t^*)}} \delta(\vec{x_{r}}-\vec{z}_{r}).
\end{equation}
Using Eqs. (\ref{j}) and (\ref{del2}), we can find the topological
charge corresponding to the current $j^{\mu}$ is the winding
number $W_{r}$. Since the topological current $j^{\mu}$ is
identically conserved, the topological charge $W_{r}$ will remain
unchanged in various evolution processes. Obviously, the above
discussion also apply to $\phi$. Therefore, at this level, the
topological evolution of the vortices for $\psi$ and $\phi$ is
relatively independent.

 At this point, a question rises:
is there any other nontrivial topological configuration formed in
the system due to the interaction between the fields $\psi$ and
$\phi$? To answer this question, let us pay attention to
coherently coupled two-component Bose-Einstein condensates studied
in Ref. \refcite{KasamatsuPRL200493}. Such a system is not
described by a gauge theory, but it has the same complex scalar
field structure with the present system. So we can use the method
presented in Ref. \refcite{KasamatsuPRL200493} to get an insight
into the topological configuration formed by the internal coherent
coupling between $\psi$ and $\phi$. To be specific, writing
$\psi=\rho\cos(\theta/2)\exp(i\varphi_1)$ and
$\phi=\rho\sin(\theta/2)\exp(i\varphi_2)$, we can introduce an
internal space vector
\begin{equation}\label{n}
\vec{n}=(\sin\theta\cos\gamma,\sin\theta\sin\gamma,\cos\theta),
\end{equation}
where $\gamma=\varphi_2-\varphi_1$. Field $\psi$ and $\phi$ tend
to their vacuum values at the space infinity. This boundary
condition compactifies space $R^2$ into $S^2$. So vector $\vec{n}$
defines a mapping: $S^2\rightarrow S^2$. Since $\pi_{2}(S^2)=Z$,
there exists another kind of topological soliton named skyrmions.
In the general nonaxisymmetric case, such a skyrmion will split
into a pair of vortex-antivortex, or a vortex molecule, which is
formed by a vortex in $\psi$ ($\phi$) and an antivortex in $\phi$
($\psi$). The concrete amounts of the bound vortex molecules and
unbound vortices in the system depend on the winding numbers of
$\psi$ and $\phi$ as well as the skyrmion numbers of the system.

Because of the existence of the vortex molecule, the detailed
evolution processes of the vortices in model (\ref{L}) is more
complicated than the vortex evolution processes in the
corresponding single scalar field model. For example, the
processes in which one vortex in a vortex molecule is replaced by
another vortex is absent in the single scalar field case. We will
leave this subject to the future studies.

\section{CONCLUSIONS}
Making use of $\phi$-mapping topological current method, we have
discussed the self-dual vortices in the Abelian Chern-Simons model
with the Chern-Simons gauge field coupling to two complex scalar
fields. This model is a kind of (2+1)-dimensional field theory
generalization of the two-condensate system, which receives wide
attention in condensed-matter
physics.\cite{BabaevPRL200288,BabaevPRL200289,JiangPRB200470,DuanPRB200674}
For each scalar field, we analytically derived an exact nontrivial
equation with a topological term which is missing in many
references. Each of these equations holds for a very wide range of
the boundary conditions. Besides, the equation relating the two
scalar fields by their topological terms was also obtained. We
calculated the angular momentum of the Chern-Simons vortices, and
the result is exactly the generalization of the angular momentum
of single scalar field Chern-Simons vortices. We also calculated
the magnetic flux of the system for different boundary conditions,
and found this magnetic flux relates the total winding numbers for
the two scalar fields in a specific way which depends on the given
boundary condition. Furthermore, we briefly discussed the
evolution processes of the vortices, and found that because of the
present of the vortex molecule, the detailed evolution processes
of the vortices in the present model is more complicated than the
vortex evolution processes in the corresponding single scalar
field model.

\section*{Acknowledgements}
It is a pleasure to thank Dr. Yong-Qiang Wang for discussions.
This work was supported by the National Natural Science Foundation
of the People's Republic of China (No. 502-041016 and No.
10705013) and the Fundamental Research Fund for Physics and
Mathematics of Lanzhou University (No. Lzu07002).

\end{document}